\documentclass[aps,prl,reprint]{revtex4-1}

\usepackage{amsmath}
\usepackage{graphicx}
\usepackage{dcolumn}
\usepackage{verbatim}

\newcommand{\rhon}[1]{\rho^{(#1)}}

\newcommand{\phins}[2]{\phi_{#2}^{*(#1)}}

\newcommand{\nuks}{\nu_{ks}}
\newcommand{\nuksr}{\nu_{ks}(\mathbf{r})}
\newcommand{\nuksnn}[1]{\nu_{ks}^{(#1)}}

\newcommand{\br}{\mathbf{r}}

\begin{document}

\title{Energy Continuity in Degenerate Density Functional Perturbation Theory}
\author{Mark C. Palenik}
\thanks{NRC Research Associate}
\email{mark.palenik.ctr@nrl.navy.mil}
\author{Brett I. Dunlap}
\affiliation{Code 6189, Chemistry Division, Naval Research
Laboratory, Washington, DC 20375, United States}

\begin{abstract}
Fractional occupation numbers can produce open-shell degeneracy in density functional theory. We develop the corresponding perturbation theory by requiring that a differentiable map connects the initial and perturbed states. The degenerate state connects to a single perturbed state which extremizes, but does not necessarily minimize or maximize, the energy with respect to occupation numbers. Using a system of three electrons in a harmonic oscillator potential, we relate the counterintuitive sign of first-order occupation numbers to eigenvalues of the electron-electron interaction Hessian.
\end{abstract}

\maketitle

When a quantum mechanical system is perturbed by a small external potential, $\lambda V^{(1)}$, it is often possible to build Taylor series in the parameter $\lambda$ connecting the eigenstates of the perturbed and unperturbed systems.  This is the premise of Rayleigh-Schr\"odinger perturbation theory (RSPT) \cite{Shavitt2009}.

In standard quantum mechanics, degeneracy means that there is not a one-to-one mapping between unperturbed and perturbed eigenvalues.  The potential $V^{(1)}$ will, in general, break the initial degeneracy, causing different linear combinations of previously degenerate states to evolve into different eigenstates.  In Kohn-Sham (KS) density functional theory (DFT) \cite{HKTheorems,Kohn1965}, on the other hand, the original degenerate state is paradoxically unique, because degenerate eigenvalues only occur for a specific set of orbital occupations.  This means that if perturbation theory can be defined, it must connect the unperturbed state to a unique perturbed state as well \cite{Cances2014}.

In standard quantum mechanics, unless we pick the correct initial linear combination of degenerate states, a discontinuous shift in the wave function is required to remain in an eigenstate after the perturbation is turned on.  This correct linear combination is determined entirely by the perturbing potential and not the initial state \cite{sakurai2011modern}.  What properties of KS DFT, then, allow for the existence of a continuous, differentiable connection between perturbed and unperturbed states where none can exist in standard quantum mechanics?

Variational minimization of the KS energy with respect to the orbitals leads to the nonlinear eigenvalue equation $H_{KS}|\phi_i\rangle=\epsilon_i|\phi_i\rangle$, where $\phi_i$ is a single-particle orbital.  The operator $H_{KS}$ is like a quantum mechanical Hamiltonian, except that it contains Coulomb and exchange-correlation (XC) potentials, which we will collectively refer to as $\nuks$.  These potentials are meant to model electron-electron interactions and introduce nonlinearity because they depend on the electron density, which in turn is determined by the orbitals.

In DFT, it is the symmetry of $H_{KS}$, which includes both the external potential and $\nuks$, that is responsible for eigenvalue degeneracy.  For an open-shell system, $\nuks$ is symmetric if and only if each element of the open shell is occupied equally.  For example, a single electron that equally occupies three \textit{p} orbitals with occupation numbers of one third will produce a spherically symmetric density and a corresponding spherically symmetric $\nuks$.

The fact that we start with equal occupation numbers is helpful, because it means that we can apply a unitary transformation that diagonalizes the first-order potential without changing the density.  However, the level splitting induced by $V^{(1)}$ means that if the occupation numbers are left unchanged, several perturbed states with different eigenvalues will be equally occupied.

\begin{figure*}[t]
	\includegraphics[width=0.655\columnwidth]{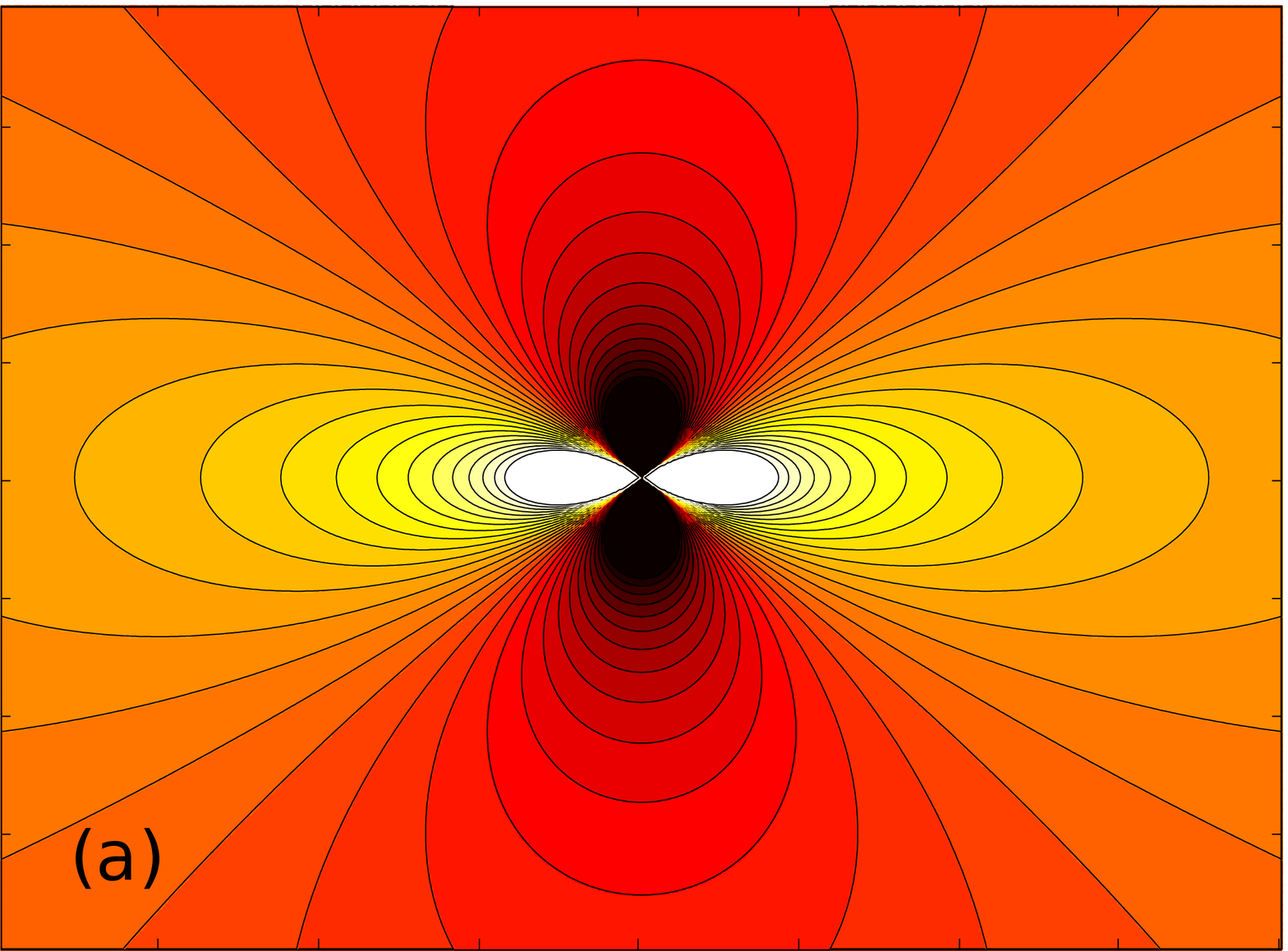}
	\includegraphics[width=0.66\columnwidth]{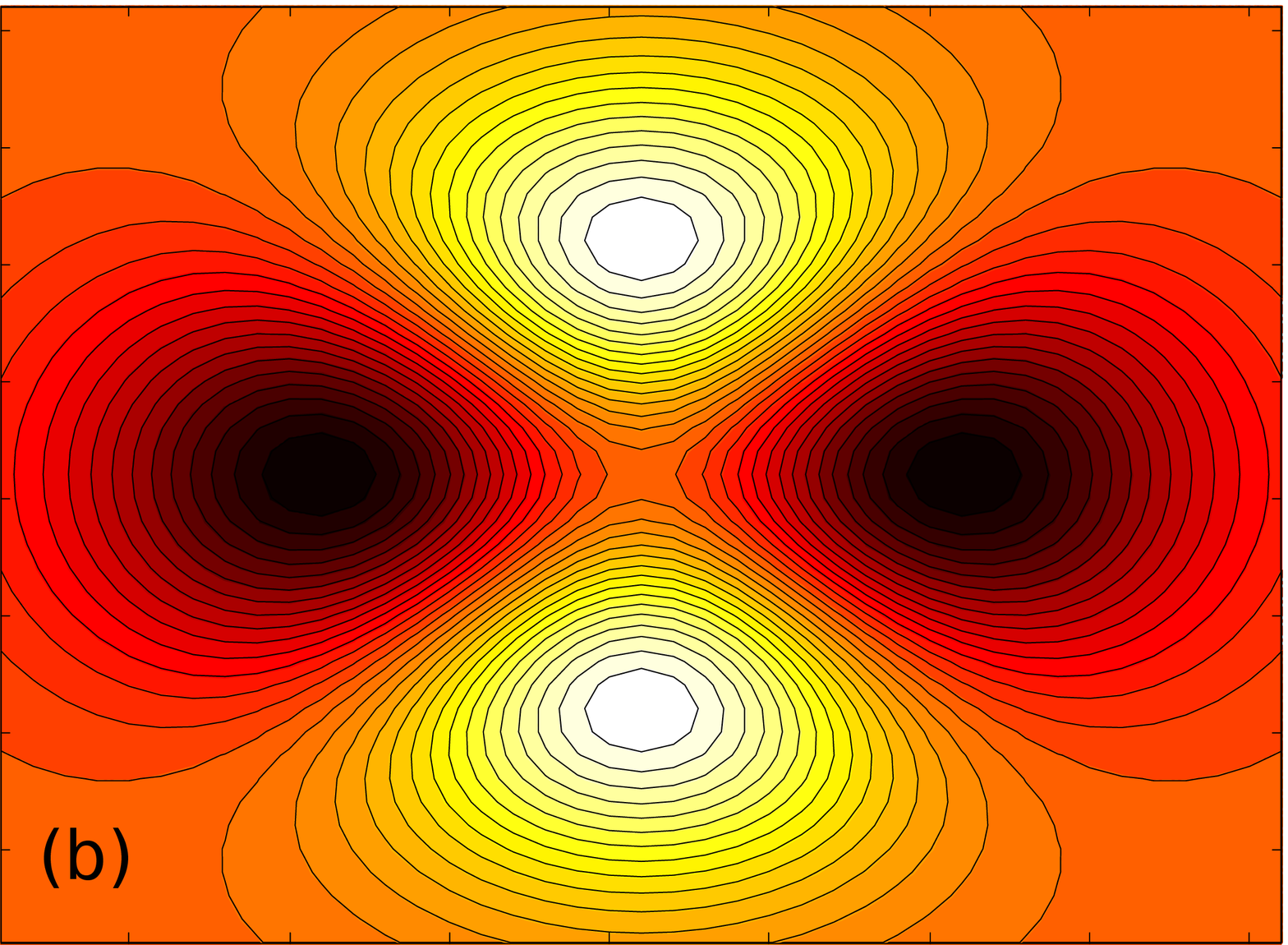}
	\includegraphics[width=0.66\columnwidth]{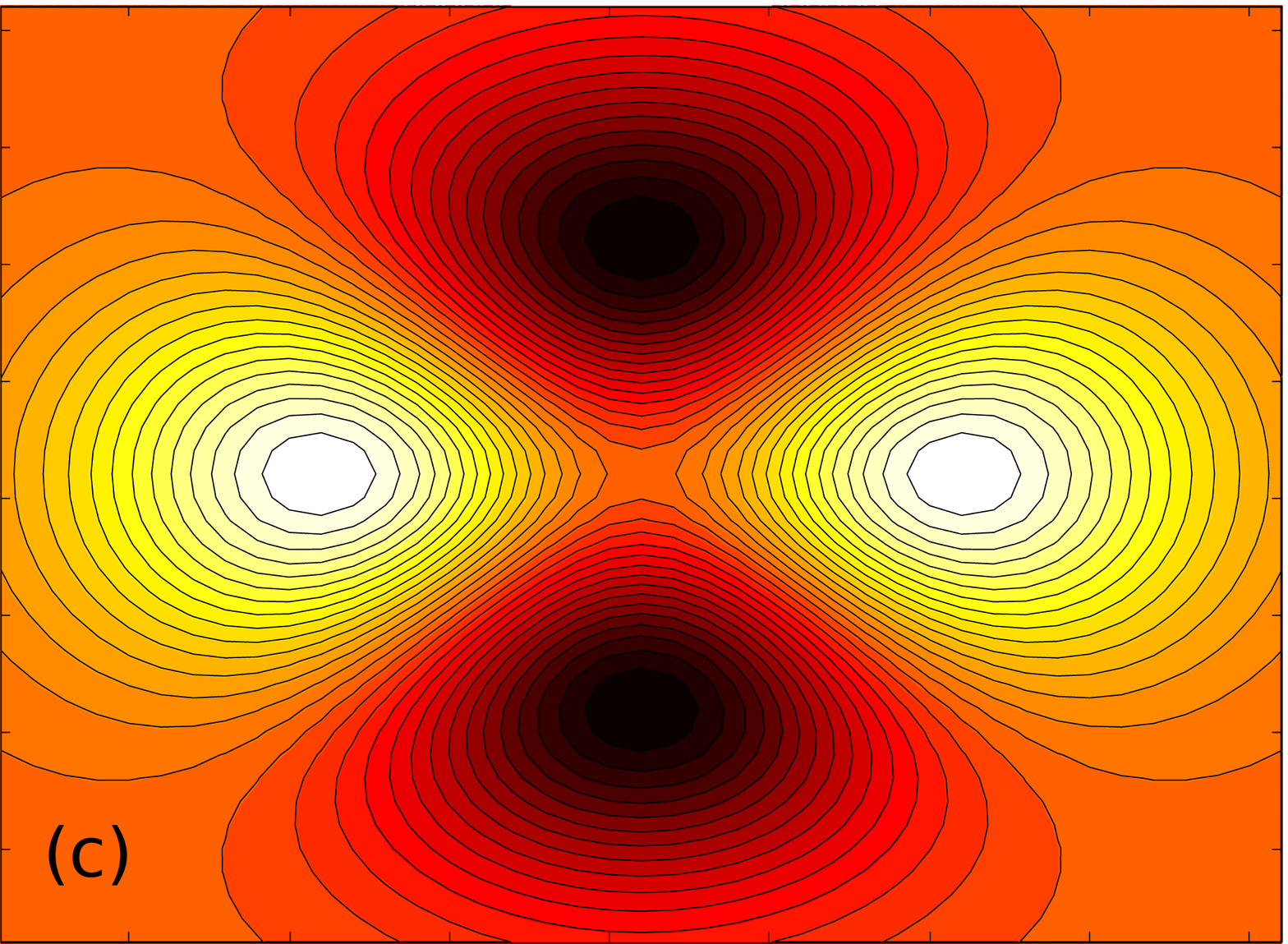}
	\caption{(a) perturbing external potential $V^{(1)}$ in the $yz$ direction (b) first-order density in the $yz$ direction for $\alpha<0.577$ without SIC (c) first-order density in the $yz$ direction for $\alpha>0.577$ or with SIC.  Lighter colors represent higher values and darker colors are lower.}
	\label{figrhoandv}
\end{figure*}

This problem can be solved by allowing the occupation numbers to change at each order.  The Coulomb and XC potentials have an order-by-order expansion, because they depend on the density, which has an order-by-order expansion.  Therefore, in DFT, it is the full first-order potential $V^{(1)}+\nuksnn{1}$ that must be diagonalized.  If the perturbation is small, transferring electrons between orbitals can change $\nuksnn{1}$ enough that the eigenvalues become degenerate again.  A mathematical proof under mild assumptions shows that two features of degenerate DFT perturbation theory are a change in natural occupation numbers at the Fermi level and a lack of eigenvalue splitting \cite{Cances2014}.

The transfer of electrons from one orbital to another does not necessarily minimize the energy but instead does whatever is required to equate the eigenvalues.  This is determined by the interdependence of each eigenvalue on all of the orbital occupations.  While one typically expects the eigenvalue of a given orbital to increase as its occupation increases, the eigenvalues also depend on the occupation of all other orbitals, and thus, electrons may end up moving between orbitals in a counterintuitive way.  For example, in Fig.~\ref{figrhoandv}(c), the occupation numbers change such that the interaction of the first-order density with the first-order potential is positive, although the total second-order energy is still negative, as one would expect from RSPT.

Before delving into this counterintuitive behavior, we will first use the requirement of differentiability to simultaneously prove that the eigenvalues remain degenerate and find equations for the orbitals at each order.  In order for a Taylor series in $\lambda$ to exist connecting the unperturbed and perturbed states, at the very least, the unperturbed state must be differentiable with respect to $\lambda$.  If we can write an explicit expression for the orbitals as a function of $\lambda$ and take its derivatives at $\lambda=0$, we can then determine the requirements for differentiability.

Such an expression can be produced by using a normalized, imaginary-time propagator, which results in an orbital with the lowest possible eigenvalue in the limit that $t\rightarrow\infty$.  Unlike in standard quantum mechanics, the choice of normalization in DFT is not arbitrary.  Although in perturbation theory, the intermediate normalization is often convenient, the Coulomb and XC functionals depend nonlinearly on the electron density, and it is therefore crucial that its magnitude, the number of electrons, is conserved.

We would like to extend this idea to the situation where multiple degenerate orbitals are fractionally occupied.  The degeneracy of the unperturbed orbitals in DFT depends on a particular choice of occupation numbers and therefore, the entire fractionally occupied state must evolve into an eigenstate, continuously as a function of $\lambda$.  An expression for a Fermi level orbital, $\phi_i$, as a function of $\lambda$, provided it overlaps the original $\phi_i$ for all values of $\lambda$, is given by
\begin{equation}
|\phi_i(\lambda)\rangle=\frac{\sum_k|\phi_k\rangle\langle\phi_k|\mathcal{T}e^{-\int_0^\infty
H'_{IP}dt}|\phi_i\rangle}{\frac{1}{N_d}\sum_m\sqrt{\sum_j|\langle\phi_j|\mathcal{T}e^{-\int_0^\infty
H'_{IP}dt}|\phi_m\rangle|^2}},
\label{PhiGround}
\end{equation}
where $H'_{IP}$ is the difference between the perturbed and unperturbed $H_{KS}$ in the interaction picture and $\mathcal{T}$ is the time ordering operator \cite{PalenikUnpublished}.  The index $m$ runs over the degenerate orbitals, and $j$ and $k$ run over all orbitals at or above the Fermi level.  $N_d$ is the number of degenerate orbitals. The factor of $1/N_d$ in the denominator comes from the fact that the degenerate orbitals are initially equally occupied and conserves the total density within the degenerate space.

The first-order orbitals are the first term in the Taylor series of $\phi_i(\lambda)$ at $\lambda=0$.  Therefore, we need to differentiate Eq.~(\ref{PhiGround}) once.  When there is no degeneracy, this reproduces the standard RSPT sum over states expression.  When there is degeneracy, the matrix elements between degenerate orbitals that appear in the derivative of Eq.~(\ref{PhiGround}) become infinite.

If perturbation theory is to give a meaningful result, all of the matrix elements involved must approach some well defined value as $t\rightarrow\infty$. Another way of stating this is to say that the time-derivative of the matrix elements must go to zero as $t\rightarrow\infty$.  Imposing this condition on the matrix elements between degenerate orbitals $\phi_k$ and $\phi_i$ at first order yields \cite{PalenikUnpublished}
\begin{equation}
\langle\phi_k|V^{(1)}+\nuksnn{1}|\phi_i\rangle=\epsilon^{(1)}\delta_{ik},
\label{EqFirstOrder}
\end{equation}
where $\epsilon^{(1)}$ is the first order eigenvalue of all of the fractionally occupied orbitals.  We can continue to higher orders by taking additional derivatives of $\phi_i(\lambda)$ at $\lambda=0$.  At each order, applying this procedure has the same effect of equating the eigenvalues.

This lack of eigenvalue splitting would, at first glance appear to cause problems for perturbation theory.  For example, the usual expression for the first-order mixing between degenerate states has first-order eigenvalue differences in the denominator.  While these are nonzero in standard quantum mechanics, they remain zero in DFT.  However, it can be shown that the $N$th-order mixing between degenerate orbitals is actually part of perturbation theory at order $N+1$ and does not affect any $N$th-order observables in DFT \cite{PalenikUnpublished}.  DFT introduces a new term through the second-order KS potential, and when this term is included, the equations can be solved without a singularity once again.  

In order to make the eigenvalues equal, we must understand how they change as electrons are transferred between orbitals.  This information can be obtained from their derivatives with respect to occupation numbers, or equivalently, the Hessian of the energy with respect to occupation numbers.  We will compute this Hessian in a way that sheds light on its relationship to the order-by-order expansion of the occupation numbers, by making a connection to our prior work on density perturbation theory \cite{Palenik2015}.  There, we showed that the electron density at order $N$ can be found directly by making the energy at order $N+M$ stationary with respect to the density at order $M$.  If we apply this same idea to fractional occupation numbers, with the constraint that the total number of electrons in the fractionally occupied orbitals is conserved, we can write the equation
\begin{equation}
\frac{dE^{(N+M)}}{dn_j^{(M)}} = \epsilon^{(N)},
\label{EqdEdn}
\end{equation}
where now, $\epsilon^{(N)}$ is a Lagrange multiplier that enforces the constraint and $n_j^{(M)}$ is the $M$th-order occupation number of the Fermi level orbital $\phi_j$. Explicitly evaluating the left hand side yields, as one would expect from Janak's theorem \cite{Janak1978}, the RSPT equation for the $N$th-order eigenvalue $\epsilon_j^{(N)}$.  The exception is that $dE^{(N)}/dn_j^{(0)}$ for $N>0$ cannot be easily evaluated, due to the self-consistent dependence of the unperturbed density on the zeroth-order occupation numbers \cite{PalenikUnpublished}.

The zeroth-order Hessian, given a fixed set of zeroth-order orbitals (as a basis for the perturbation expansion), is the second derivative of $E^{(N+M)}$ with respect to $n_j^{(M)}$ and $n_k^{(N)}$, given by
\begin{equation}
\frac{d^2E^{(N+M)}}{dn_j^{(M)}dn_k^{(N)}}=\frac{d\epsilon_j^{(N)}}{dn^{(N)}_k} = \int \rhon{0}_j(\br)\frac{\delta\nuksr}{\delta\rho(\br')}\rhon{0}_k(\br')d\br d\br',
\label{EqEHess2}
\end{equation}
where we make the definition $\rho_j^{(0)}(\br)=\phi_j^*(\br)\phi_j(\br)$.  The term $\delta\nuksr/\delta\rho(\br')$ is the Hessian of the electron-electron interaction energy and comes from the first-order KS potential \cite{Palenik2015}.  This Hessian is neither positive definite nor negative definite, due to the positive contribution of the Coulomb energy and the negative contribution of XC \cite{Dunlap2016}. Therefore, the extremum we find, in general, will be an energy saddle point.

Because the Hessian is equal to the derivative of the $N$th-order eigenvalues with respect to the $N$th-order occupation numbers, at all orders, this matrix must be inverted to find the corresponding occupation numbers.  If we know the basis that diagonalizes the entire first-order potential, in this basis, we could rearrange Eq.~(\ref{EqFirstOrder}) to get
\begin{equation}
\begin{split}
&\sum_jn_j^{(1)}\langle\phi_i|\int\frac{\delta\nuksr}{\delta\rho(\br')}\rho_j^{(0)}(\br')d\br'|\phi_i\rangle
=\epsilon^{(1)}\\
&-\langle\phi_i|V^{(1)}+2Re\sum_j\int\frac{\delta\nuksr}{\delta\rho(\br')}\phi_j^{*(1)}(\br')\phi_j(\br')d\br'|\phi_i\rangle.
\end{split}
\label{EqN1LHS}
\end{equation}
The left hand side is the zeroth-order Hessian times the first-order occupation numbers.  This matrix determines the manner in which the initial occupation numbers extremize the unperturbed energy, assuming the unperturbed orbitals are unchanged for a small change in occupation numbers, and also the way the $M$ and $N$th-order occupation numbers extremize the $M+N$th-order energy.  The negative contribution of XC means that the unperturbed state is not necessarily an energy minimum.  This has a profound effect on the behavior of the perturbed occupation numbers.

\begin{figure*}[t]
	\includegraphics[width=\columnwidth]{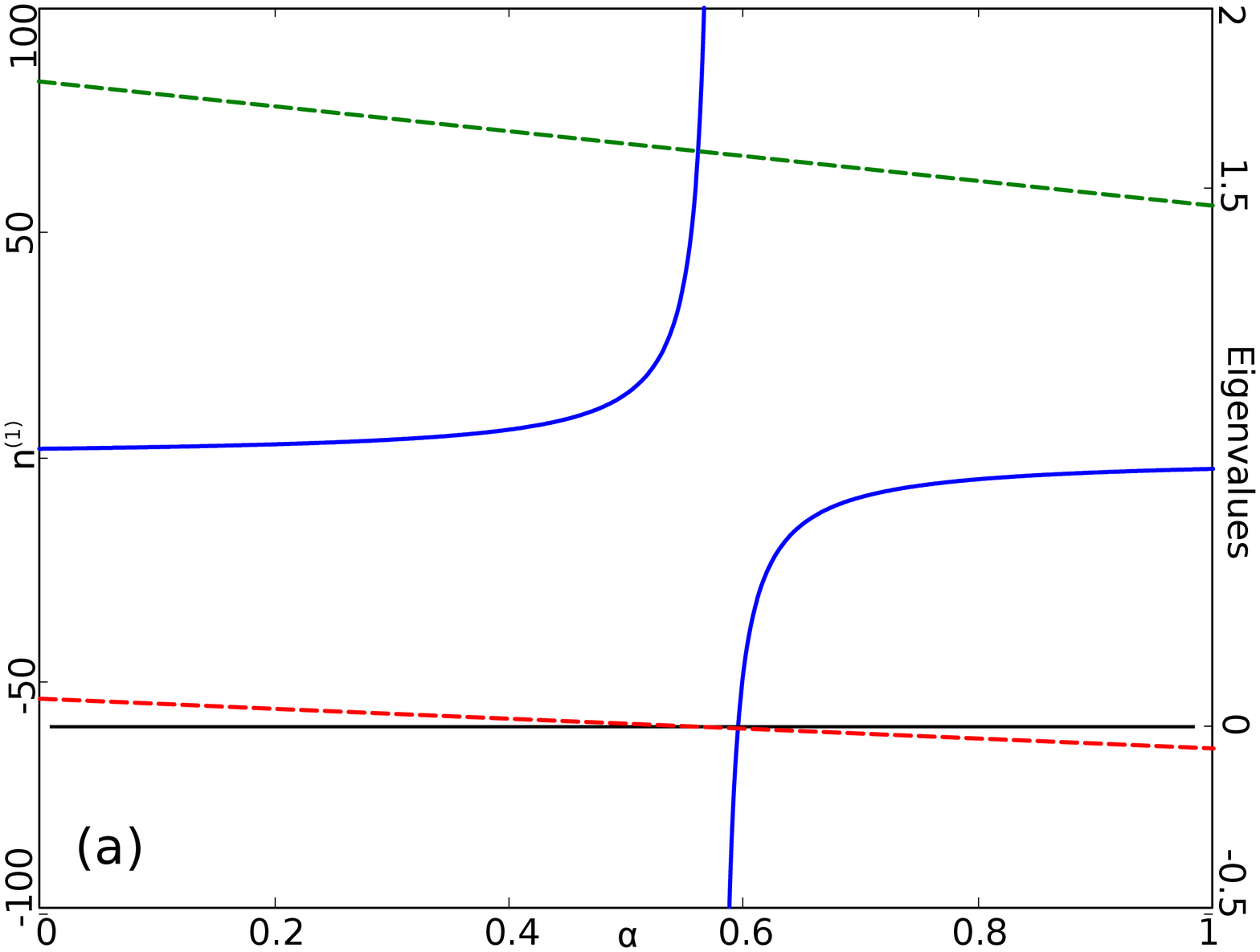}
	\includegraphics[width=\columnwidth]{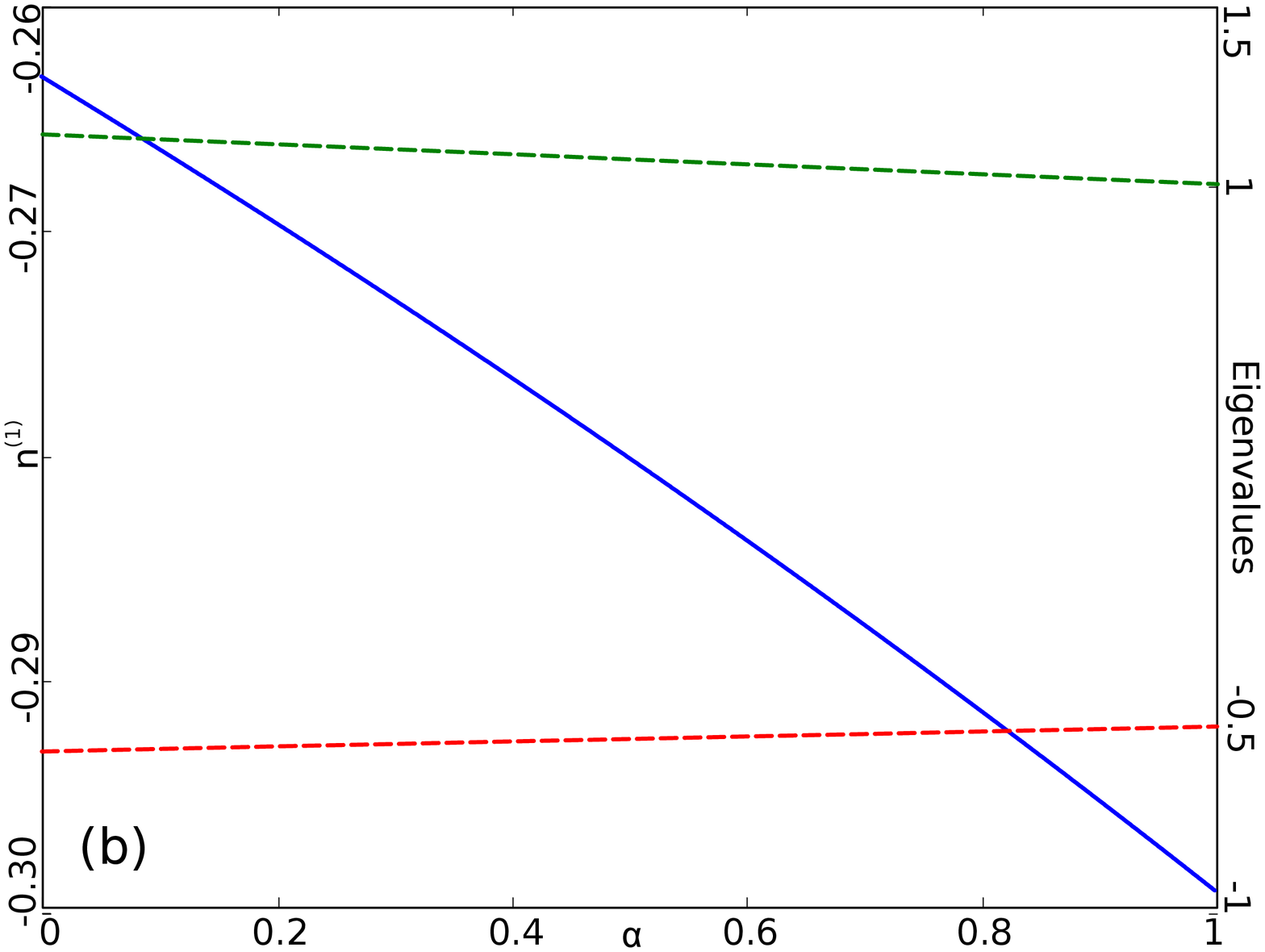}
	\caption{First order change in occupation numbers when $\omega Q=1$ (solid lines) and Hessian eigenvalues (dashed lines) versus the parameter $\alpha$.  (a) without SIC (b) with SIC.  Note the vastly different scales on the vertical axis for $n^{(1)}$.  The horizontal black line in (a) highlights the eigenvalue zero-crossing.  The lower Hessian eigenvalue is doubly degenerate.}
	\label{fignvsalpha}
\end{figure*}

We can demonstrate this by looking at a model problem that can be solved analytically.  We will apply a perturbing electric quadrupole potential, given by $V^{(1)}(\br) = Q(y^2-z^2)/|\br|^5$ to a system of three electrons in a harmonic oscillator potential, using the X$\alpha$ functional for exchange and correlation, with and without a self-interaction correction (SIC) \cite{Perdew1981}.  The contribution of XC is scaled by the parameter $\alpha$.

Two electrons occupy the lowest state with opposite spins, and the third equally occupies all three degenerate first-excited states.  We solve this system in the limit that the oscillator frequency, $\omega$, is infinite.  This makes the unperturbed ground state independent of $\alpha$ because the interaction with the external potential dominates, and it allows us to neglect mixing with virtual orbitals.

Because there is no mixing with virtual orbitals, the term on the right hand side of Eq.~(\ref{EqN1LHS}) involving $\phins{1}{j}\phi_j$, when summed over all $j$, is zero.  If desired, we can add the SIC by multiplying the left hand side by $1-\delta_{ij}$.  As Perdew and Zunger state, this SIC does not truly remove all self-interactions when fractional occupation numbers are used.  If self interactions were completely removed, the entire left hand side of Eq.~(\ref{EqN1LHS}) would be zero because there is only a single electron within the degenerate space.  Although it is clear that for a single electron, the left hand side of Eq.~(\ref{EqN1LHS}) is zero in a theory completely free of self-interactions, it is much less clear what the correction should be for two or more electrons.

The first-order occupation numbers can be used to find the first through third-order energies \cite{Wigner1935,Angyan2009,Cances2014}.  The first and third order energies, in this case, are zero.  The second-order energy is negative and given by
\begin{equation}
E^{(2)} = -\frac{1}{2}\sum_{ij} n_i^{(1)}n_j^{(1)}\int \rhon{0}_i(\br)\frac{\delta\nuks(\br)}{\delta\rho(\br')}\rhon{0}_j(\br')d\br.
\end{equation}
The terms that explicitly depend on $V^{(1)}$ either cancel from $E^{(2)}$ or are zero because there is no mixing with virtual orbitals \cite{PalenikUnpublished}.  This makes the second-order energy negative, even in a situation like the one depicted in Fig.~\ref{figrhoandv}(c), where the interaction of the first-order density with the first-order potential is positive.



Performing the diagonalization required by Eq.~(\ref{EqFirstOrder}) is simple in this problem, in part because the first excited states of the Harmonic oscillator have odd parity, meaning that the $\nuksnn{1}$ term is diagonal in any basis.  Diagonalizing $V^{(1)}$ only requires aligning the the Hermite polynomials in the first excited states of the harmonic oscillator with the axes of the quadrupole.  Solving for $n^{(1)}_j$ is then a matter of inverting the Hessian from Eq.~(\ref{EqEHess2}).  The matrix elements are proportional to $\sqrt{\omega}$ (and the exchange portion also has a factor of $\alpha$), while the matrix elements of $V^{(1)}$ are proportional to $Q\omega^{3/2}$. The $\langle\phi_x|V^{(1)}|\phi_x\rangle$ and $\langle\phi_z|V^{(1)}|\phi_z\rangle$ matrix elements have opposite signs, while $\langle\phi_z|V^{(1)}|\phi_z\rangle$ is zero, and so, the same is true for $n^{(1)}_x$, $n^{(1)}_y$, and $n^{(1)}_z$.  Therefore, we can specify the first-order occupation numbers by a single parameter, $n^{(1)}$, which we will take to be $n^{(1)}_z$.

In Fig.~(\ref{fignvsalpha}), we have plotted $n^{(1)}$ alongside the Hessian eigenvalues as a function of $\alpha$.  Two of the three eigenvalues are degenerate, represented by the lower line in both (a) and (b).  Without the SIC, at $\alpha=0.577$, the off-diagonal and diagonal elements of the Hessian become equal, causing two of the three eigenvalues to go to zero.

When $\alpha$ is greater than $0.577$, these two eigenvalues become negative, meaning that the energy is extremized to a saddle point.  Here, $n^{(1)}$ also becomes negative, which means that electrons move from $\phi_z$ into $\phi_y$ [Fig.~\ref{figrhoandv}(c)].  The perturbing potential and first-order density are shown in Fig.~\ref{figrhoandv} along the $y$ (horizontal) and $z$ (vertical) axes.  

Adding a SIC zeros the diagonal elements of the Hessian, which then has two negative eigenvalues over the entire range of $\alpha$ between zero and one [Fig.~\ref{fignvsalpha}(b)].  All three eigenvalues change sign at $\alpha=8.272$, well outside of the physically reasonable range of $0.6$ to $1.0$.  This occurs when the off-diagonal elements, and therefore, the entire matrix become zero.

With the SIC, we can always lower the energy by moving electrons into a single orbital.  This removes the electron-electron interaction energy, which is dominated by the positive Coulomb term.  Therefore, it is obvious that the fractionally occupied state is not an energy minimum with respect to occupation numbers.  Similarly, without the SIC, the Hessian is not positive definite when $\alpha>0.577$.  Negative Hessian eigenvalues mean that there is some combination of electron transfers between degenerate orbitals that can lower the total energy.

The occupation numbers are proportional to the inverse of this same Hessian at all orders.  The presence of negative eigenvalues can change the direction of electron transfer, causing the electron-density to behave in counterintuitive ways, such as in Fig.~\ref{figrhoandv}(c).  The eigenvalues of the degenerate orbitals are determined by the interaction of electrons with the entire KS potential, which includes Coulomb and XC portions, and perturbation theory will cause electrons to rearrange themselves in whatever way is necessary to maintain degeneracy after the external perturbing potential is applied.



\begin{acknowledgments}
This work is supported by the Office of Naval Research, directly and through the Naval Research Laboratory. M.C.P. gratefully acknowledges an NRC/NRL Postdoctoral Research Associateship.
\end{acknowledgments}

\bibliography{citations}{}
\end{document}